# Decoding the Hot-Mitochondrion Paradox

Peyman Fahimi,*[(a)] Michael Lynch,[(b)] Chérif F. Matta,*[(c-e)]

[(a)] *Department of Mathematics & Statistics, Dalhousie University, Halifax, NS B3H4R2 Canada.* [(b)] *Center for Mechanisms of Evolution, Biodesign Institute, Arizona State University, Tempe, AZ 85287 USA.* [(c)] *Department of Chemistry and Physics, Mount Saint Vincent University, Halifax, NS B3M2J6 Canada.* [(d)] *Dép. de chimie, Université Laval, Québec, QC G1V0A6 Canada.* [(e)] *Department of Chemistry, Saint Mary's University, Halifax, NS B3H3C3 Canada.*

* *E-mails: fahimi@dal.ca; cherif.matta@msvu.ca*

## Abstract

In a 2018 paper and a subsequent article published in 2023, researchers reported that mitochondria maintain temperatures 10–15 °C higher than the surrounding cytoplasm - a finding that deviates by 5 to 6 orders of magnitude from theoretical predictions based on Fourier's law of heat conduction. In 2022, we proposed a solution to this apparent paradox. In the present perspective, we build upon that framework and introduce new ideas to further unravel how a biological membrane - whether of an organelle or a whole cell - can become significantly warmer than its environment. We propose that proteins embedded in the inner mitochondrial membrane (IMM) can be modeled as ratchet engines, introducing a novel, previously overlooked mode of heat transfer. This mechanism, coupled with localized heat release during the cyclical dehydration-translocation-hydration of ions through membrane proteins, may generate transient but substantial temperature spikes. The cumulative thermal occupancy of these microscopic events across the three-dimensional surface of the IMM can account for the elevated temperatures detected by molecular probes.

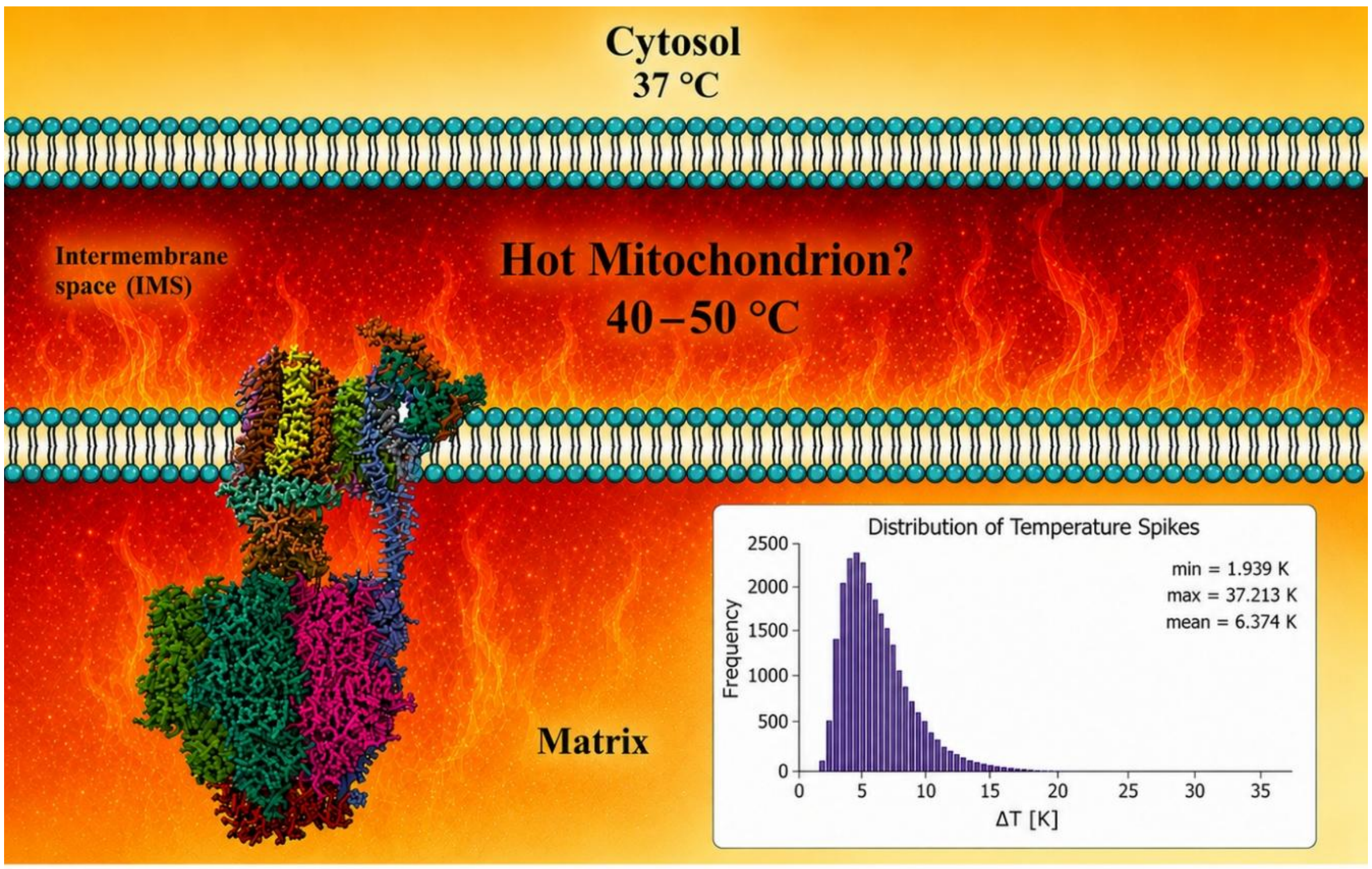

**TOC Graphic: A distributed mechanism of mitochondrial heat generation is proposed in which all ion-translocating proteins embedded within the inner mitochondrial membrane contribute to localized heat generation.**
All proteins embedded within the inner mitochondrial membrane (IMM) – with ATP synthase shown here as an example – are proposed to function as ratchet engines. Concurrently, the cyclical dehydration-translocation-hydration processes of ions moving between the intermembrane space (IMS) and the mitochondrial matrix act as distributed heat-release events across countless segments of the three-dimensional IMM, collectively shaping the temperature gradient between the mitochondrion and the surrounding cytoplasm. The total cumulative thermal occupancy is defined as the product of the total ion flux generated by all ion-translocating proteins and the effective lifetime of the corresponding thermal spikes.

## 1. Introduction

Fluorescence-based intracellular thermometry has led to the provocative suggestion that mitochondria can operate 10-15 °C hotter than their surrounding cytoplasm [1,2]. This surprising claim appears to contradict classical steady-state heat-transfer estimates based on Fourier's law by 5-6 orders of magnitude, a discrepancy we refer to as the "hot-mitochondrion paradox" (HMP)[1]. A brief plain-language overview of these experimental claims and their potential biological implications is provided in Box 1. The discovery of temperature gradients within living cells has significant implications for human health and disease. For instance, thermoresponsive nanocarriers can be utilized to deliver anticancer drugs specifically to mitochondria in cancer cells [3]. Furthermore, mitochondrial temperature may serve as a biomarker for hepatocellular carcinoma [4].

[1] A table defining abbreviations and another defining the symbols appearing in this paper are available in the Supporting Information.

**BOX 1. Plain-language summary of the Hot-Mitochondrion Paradox**

In 2018, an intriguing article appeared claiming that mitochondria operate at a much higher temperatures than the surrounding cytoplasm, and that mitochondria are, in fact, "hot". These claims are based on interpreting fluorescence intensity measurements of a dye that is known to concentrate inside mitochondria. The paper in question is that of Chrétien *et al.* [1], which was followed-up in 2023 with another paper [2] by the same group confirming their results that mitochondria operate at a temperature ~10ºC – 15ºC higher than their surrounding temperature and extended their results to other experimental model systems. In the very issue of *PLOS Biology* where the 2018 paper appears, a guest-editorial by Nick Lane expressed skepticism regarding claims of a "hot mitochondrion," while not ruling it out entirely [5]. These claims have sparked a heated debate in the literature where, on one side, Chrétien *et al.*'s results were never falsified and still stand unchallenged while steady-state models of heat transfer place a maximal temperature differential at $10^{-5}$ K across dimensions typical of living cells [6–8]. This has been termed the $10^5$ gap in the literature and remained a paradox until a tentative resolution appeared in 2022 [9]. One of the goals of this article is to simplify this proposed resolution in plain language and to elaborate and extend these ideas.

A variety of experimental approaches have since been developed to probe subcellular temperature, including small-molecule dyes, genetically encoded sensors, and nanoparticle-based probes. These methods yield a mixed picture: some studies report mitochondrial or nuclear regions that appear several degrees warmer than the surrounding cytoplasm, whereas others find temperature differences of <1 °C or no detectable temperature difference at all. The underlying measurement principles, calibration strategies, and known pitfalls of these techniques are summarized in Box 2.

**BOX 2. How intracellular temperature is measured: techniques and pitfalls**

Among the early experiments, Okabe *et al.* [10] reported that the temperature of the nucleus and centrosome in COS7 cells is approximately one degree Celsius higher than that of the surrounding cytoplasm. In 2018, Chrétien *et al.* [1,2] proposed that mitochondria operate at significantly higher temperatures than traditionally assumed, based on experiments using a molecular thermometer dye called mito-thermo-yellow (MTY) [11]. This fluorescent probe predominantly localizes in the mitochondrial inner membrane and its matrix-facing side [1,12]. Contrary to the long-held belief that mitochondrial temperature aligns with the ambient body temperature of 37°C, Chrétien *et al.* suggested that mitochondrial temperatures in human cells could reach up to 50°C. Given that mitochondria are particulate, strategically positioned, and serve as the central hub for ATP production and heat generation (via proton leak [13–16]), it is unsurprising that they are hotter than the average body temperature. However, the real debate is how much hotter.

The extraordinary finding of Chrétien, Rustin, and co-workers has sparked vigorous debate and opened up new avenues for research [5,17]. The problem is that a temperature difference ($\Delta T$) of the order of 10ºC over distances of a few nanometers implies astronomically elevated temperature gradients that cannot be maintained with any conceivable material. Nevertheless, the results of Chrétien *et al.* [1,2,12] are reproducible and these workers appear to have taken every possible precaution to mitigate interferences such as the effect of pH or dye concentration on the fluorescence quenching. However, Arai *et al.* [11] emphasized that while factors like pH, viscosity, metal ions, and oxygen species have been tested on MTY in cuvette experiments, these tests do not fully represent the complex molecular environment that MTY experiences inside cells, which differs significantly from simple aqueous buffer conditions.

Chrétien *et al.* [1,2,18] obtained their results in human embryonic kidney 293 cells and primary skin fibroblasts with the electron transport chain fully active under various physiological conditions. The conclusions were drawn from observed MTY quenching intensity as a function of temperature, validated through ***in vitro*** calibration curves. These researchers invested significant effort in ruling out potential artifacts or interference that might produce false positives. For instance, they demonstrated that the fluorescence decrease was unaffected by either dye concentration or local pH - two critical variables since mitochondrial dye concentrations are difficult to precisely measure and pH across the inner membrane can vary by up to one unit. According to the calibration curve, as also reported in other studies [11], the MTY fluorescence intensity in aqueous solutions decreases by 2.7% per degree Celsius of temperature elevation, independent of pH, oxygen concentration, or ionic strength.

Other researchers have developed distinct methods to map intracellular temperatures across mitochondria, nuclei, lysosomes, and other localized intracellular compartments, obtaining results consistent with those of Chretien *et al*. [1] in mammalian cell lines and yeast cells [19], murine bladder cancer MB49 cells [20], breast cancer MDA-MB468 cells [21], patient-derived tumor samples [22], HeLa cells [23], mouse brains [24], mammalian and insect cell lines [2], hippocampal neurons under modulated firing activity [25], and calcium-induced neurons [26].

Some authors, however, have raised concerns about the reliability of these experiments. For example, real-time temperature mapping of fixed A549 cells (human alveolar basal epithelial adenocarcinoma cells) indicates that localized mitochondrial heating results in a temperature difference of less than 1°C within the cell [27]. Similarly, temperature variations of less than 1°C in the nucleus and cytoplasm of live HeLa cells have been reported [28]. Other experiments have revealed that respiratory complex I becomes unstable at temperatures exceeding 43°C [29]. Concerns have also been raised regarding potential biases in measurements obtained with green fluorescent protein (GFP) nano-thermometers and semiconductor nanocrystals, which require careful consideration [30]. A recent study using nanodiamond nanothermometry reported that metabolic stimulation does not cause any temperature change in macrophages [31]. The authors noted that changes in the electrical field on the surface of the nanodiamond could be misinterpreted as temperature variations [31]. In addition, Treberg and Mailloux [32] recently argued that the kinetics of MTY fluorescence are inconsistent with the expected dynamics of mitochondrial heat production and dissipation, suggesting that MTY fluorescence may not uniquely report intramitochondrial temperature (see also the accompanying editorial commentary [33]).

A temperature difference of 10–15°C across just a few nanometers of the inner mitochondrial membrane may increase the production of reactive oxygen species (ROS) [34] and proton leak [34], alter membrane structure (including the loss of cristae folds [35] and changes in overall spontaneous curvature [36]), affect lipid phase transitions and fluidity (which are highly dependent on the presence or absence of cardiolipins [37,38]), and destabilize associated proteins [39] - all contributing to mitochondrial dysfunction. How does the mitochondrion modulate or buffer these effects?

A thermal stability atlas of 43,000 proteins spanning 13 species, from archaea to humans, reveals that protein melting temperatures range from 30°C to 90°C [39]. Respiratory chain proteins are predominantly stable across species, with human mitochondrial proteins typically functioning at 46°C [39]. A bioinformatics study suggests that the high expression of heat shock proteins (Hsps) in mitochondria plays a crucial role in safeguarding critical macromolecules from melting and mitigating the temperature-induced increase in ROS production [40]. If true, such a significantly higher temperature at which mitochondria operate would require some revision of textbook kinetics and thermodynamics of biochemical reactions happening therein [17].

Going back to the report of Chrétien *et al.* [1,2], a 10ºC - 15ºC temperature difference across the thickness of the mitochondrial membrane would imply an extraordinarily steep temperature gradient. Unsurprisingly, their claim has been met with skepticism. That skepticism is based on steady-state heat transfer considerations that appear inconsistent by a staggering 5 to 6 orders of magnitude [5–8]. The discrepancy is so substantial that it has acquired the designation of the "$10^5$ gap" between experimental claims and theoretical predictions [5,6] (Box 3).

**BOX 3. Why Fourier's law fails: the origin of the $10^5$ gap**

Fourier's law of heat conduction describes how heat flows from warmer to cooler regions. In its scalar steady-state form, the rate of heat production $\dot{\boldsymbol{Q}}$ (in units of energy per unit time, i.e., watts or joules per second) depends on three factors: the temperature difference $\Delta\boldsymbol{T}$ between two regions, the material's thermal conductivity $\boldsymbol{\kappa}$, which quantifies how easily heat flows through it, and the characteristic length $\boldsymbol{L}$ over which the heat is conducted. In a study by Baffou *et al.* [6], steady-state estimates based on Fourier's law ($\dot{\boldsymbol{Q}} = -\boldsymbol{\kappa}\boldsymbol{L}\Delta\boldsymbol{T}$) suggest that a mammalian cell, modeled as a spherical heat source with a linear dimension of 10 μm and assuming uniform heat generation throughout its volume (i.e., without localized sources such as mitochondria), producing 100 pW of thermal power in a watery environment with $\boldsymbol{\kappa}$ = 1 $W \cdot m^{-1} \cdot K^{-1}$, would result in a maximum temperature increase of approximately $10^{-5}$ K relative to its surroundings [6]. Macherel and colleagues [8] corroborated these findings by examining heat transfer mechanisms - conduction, convection, and radiation - and modeling the temperature distribution within an idealized spherical mitochondrion. In a recent study [41], the authors extended the conventional steady-state Fourier heat-diffusion framework by considering a hypothetical nonequilibrium scenario in which the inner mitochondrial membrane acts as a reverse heat engine (i.e., a heat pump). In this model, metabolic work is used to transfer thermal energy against the temperature gradient, from cooler regions to warmer regions [41]. Using the Second Law of Thermodynamics, they derived an upper bound on the temperature difference that could be maintained between the mitochondrial matrix and the surrounding cytoplasm. Even under idealized conditions, they found that the maximum temperature difference would be only about $5 \times 10^{-3}$ K [41].

How can the rate of heat production per cell be estimated? In heterotrophic eukaryotic cells, the primary source of heat production is mitochondrial proton leak** - both basal and inducible [13,14,16,42]. Proton leak refers to the movement of protons from the intermembrane space (IMS) back into the mitochondrial matrix via water wires (WWs), adenine nucleotide translocases (ANTs), and uncoupling proteins (UCPs), bypassing ATP synthase. This process dissipates the Gibbs free energy ($\Delta G$) - the usable energy available to do cellular work - of proton translocation across the IMM as heat. Gibbs free energy, given by the equation $\Delta G = \Delta H - T\Delta S$, reflects the balance between energy released (enthalpy, $\Delta H$) and energy lost to a state of disorder (entropy, $\Delta S$). In mitochondria, this energy is normally used to drive ATP synthesis by moving protons against a pH gradient and electric potential. In this context, the entropy term ($T\Delta S$), where $T$ is the absolute temperature and $S$ is entropy, is negligible compared to the enthalpy term ($\Delta H$) [43]. Therefore, during proton leak, Gibbs free energy and heat can be used interchangeably. Basal proton leak alone accounts for at least 20% of a cell's standard metabolic rate (MR) [13,42]. Therefore, to estimate the rate of heat production per cell, one needs the total metabolic rate - approximately 90% of which is due to mitochondrial oxidative phosphorylation in aerobic cells - and multiply that by the fraction lost as heat through proton leak (i.e., MR × 90% × 20%).

Baffou *et al.* [6] considered the entire cell as a heat source and estimated a temperature increase of only $10^{-5}$ K, revealing a discrepancy of at least $10^{5}$-fold between theory and experimental observations. This issue is further accentuated by the significantly smaller size of mitochondria, which range from 0.5 to 1 μm, and the metabolic rate per mitochondrion [44,45], estimated to be approximately $3 \times 10^{-4}$ pW in actively growing *Tintinnopsis vasculum* (a ciliate) and up to 1.5 pW in actively growing *Saccamoeba limax* (an amoeba), based on their cellular metabolic rates [46] and mitochondrial populations [47].

** Recently, Namari *et al.* [48] proposed a complementary mechanism in which a substantial fraction of mitochondrial heat production arises from overpotential-driven energy dissipation during electron transfer reactions in the respiratory chain, particularly at Complex IV. Overpotential is the excess driving force beyond the equilibrium redox potential required to drive electron transfer reactions under non-equilibrium conditions. The authors estimate the total redox potential difference (equivalently, the total Gibbs free energy change expressed as a potential) across the electron transport chain to be approximately 1.04 V, of which about 65% is predicted to be dissipated as heat. However, even this recently reported thermogenic contribution - which is considered substantial relative to proton leak - is still far smaller than the transient heat-release events discussed in our paper (see Table 1), which arise from ion hydration (or protonation–hydration in the case of protons). As an example, based on the values reported by the authors, a maximum of 380 kJ $mol^{-1}$ $(O_2)^{-1}$ is dissipated as heat through overpotential losses. The authors further estimate that, at a membrane potential of 0.15 V, approximately 13–18

protons are pumped into the intermembrane space per $O_2$ reduced. This corresponds to only about 20–30 kJ mol$^{-1}$ ($H^+$)$^{-1}$ dissipated as heat through overpotential during electron transport. In contrast, the Gibbs free energy released upon protonation and hydration of protons after their translocation into the intermembrane space is reported in Table 1 to be 1183 kJ mol$^{-1}$ ($H^+$)$^{-1}$. Therefore, on a per-proton basis, the heat released by protonation–hydration is at least ~40-fold greater than the heat dissipated through overpotential during electron transport chain reactions.

In the introduction of his book titled "*Paradoxes*", Sainsbury defines a paradox as: "*an apparently unacceptable conclusion derived by apparently acceptable reasoning from apparently acceptable premises*." [49] (See also Refs: [50,51]). Given the 1-million-fold disagreement between experiment and steady-state theory predictions, we have a paradox whereby the expectation of theory is in total disagreement with experimental observations. What has gone wrong? Is the interpretation of the experimental measurements at fault? Is the theoretical modeling applied outside of its limits of validity?

The present authors termed this situation "*the hot-mitochondrion paradox*" (HMP) and proposed a possible resolution [9]. The apparent paradox is interpreted as arising from the application of steady-state heat-transfer concepts to a system dominated by localized transient heat-release events and protein-scale thermal transport. The central thesis of this perspective is that both effects must be considered simultaneously when interpreting intracellular thermometry measurements. In the original paper, the proposed resolution of the paradox is rather technical, so we here provide a qualitative and intuitive explanation of its principal points. Building upon our previous theory, we take it a step further, offering a perspective on the potential origins of hot membranes in organelle or cell membranes.

## 2. A resolution of the "hot-mitochondrion paradox" (HMP)

Fourier's law of heat transfer rests on two main pillars: (1) the rate of heat production, and (2) thermal conductance.

1. As briefly discussed in the introduction, the primary source of heat in heterotrophic eukaryotic cells arises from mitochondrial proton leak. This is a long-lived, steady source of heat that dominates the average thermal output of the cell over extended timescales. However, transient heat bursts - much larger in magnitude - occur during ion translocation events[2]. These are typically not counted in the bulk heat production because they average out over time due to the interplay of endergonic reactions in the IMS and exergonic reactions in the mitochondrial matrix (and vice versa), as demonstrated in our previous work [9].

2. The inner mitochondrial membrane (IMM) is highly selective, permitting only

[2] Throughout this study, the term "ion-translocating proteins" is used in a broad physicochemical sense to denote inner mitochondrial membrane proteins that transport electrically charged species, including both inorganic ions (e.g., $H^+$, $Ca^{2+}$, $Na^+$, $K^+$, and $Mg^{2+}$) and charged metabolites, nucleotides, and cofactors (e.g., ATP, ADP, phosphate, citrate, malate, glutamate, and other substrates).

specific ions to traverse through designated proteins or enzymes. This selectivity reflects the membrane's relative resistance to ion flow and thus limits free diffusion of ions between the matrix and IMS. While macroscopic heat transfer is commonly understood to occur via conduction, convection, radiation, and advection, the mechanisms by which heat is transferred through single proteins during ion translocation remain poorly understood. Numerous types of proteins have been modeled as ratchet engines in the literature [52–60] - a concept we will elaborate on in Box 4. In our previous study [9], we demonstrated that modeling ATP synthase as a ratchet engine (or axle-vane engine) results in exceptionally low thermal conductance, implying that ATP synthase conducts heat very weakly (this does not imply that all ratchets have low thermal conductance; the effect is protein-specific; see Box 4). The ratchet-engine framework is not introduced as a mechanism of heat generation. Rather, its role is to provide a microscopic description of thermal conductance at the level of individual membrane proteins. In the present perspective, transient heat-release events associated with ion translocation provide the heat source, whereas the ratchet-engine framework provides the transport model governing how that heat is exchanged across the membrane. In the present work, we generalize this framework to all IMM-embedded proteins and address the following key questions:

a) What is a ratchet engine, and which proteins in the IMM can be modeled as such?
b) How are local heat differentials generated by different ions during translocation through the IMM?
c) What is the correct interpretation of the signals recorded by molecular probes?

**BOX 4. The ratchet-engine model explained intuitively**

A ratchet engine, by definition, consists of a vane and a ratchet enclosed in two separate, thermally isolated boxes filled with gas at two distinct temperatures, $T_1$ and $T_2$. These boxes are mechanically connected via a thermally insulated axle** [53,61]. Thus, a ratchet engine involves at least two forms of asymmetry and imbalance: one thermal (i.e., a temperature difference) and the other mechanical or geometric. For simplicity, following the approach developed by Parrondo and Español [62], we consider a model with two vanes instead of a vane and a ratchet. We refer to this simplified setup as the axle-vane engine, as illustrated in Figure 1.

This engine operates unidirectionally under non-equilibrium conditions because the temperature - and thus the average kinetic energy of the gas molecules - is higher in one box than in the other (Figure 1). As a result, the Brownian motion of gas particles is more vigorous in the hotter box, leading to an imbalance in random collisions with the vanes. This asymmetry generates a net force that induces unidirectional motion via the rotation of the first vane. This rotation is mechanically transferred to the second vane in the other box via the axle. Consequently, the kinetic energy of the gas molecules in the second box increases, indicating that heat has been transferred mechanically despite thermal isolation. In other words, *heat transfer is not mediated by molecular collisions and vibrations but rather occurs through the classical mechanical transmission of*

*Brownian motion* - facilitated by the relative (unidirectional) angular fluctuations of vanes immersed in different thermal baths and connected by the axle.

Note that this represents a novel mode of heat transfer, distinct from the well-known mechanisms such as conduction, convection, and radiation. Indeed, the macroscopic view suggests that thermal conductance depends on the constitutive material of the system, which determines how easily heat flows from hot to cold regions. In contrast, the microscopic view provided by the ratchet-engine model emphasizes how the mechanical properties of a single molecule - together with the friction between that molecule and its surrounding environment - shape the thermal conductance. Thus, the ratchet-engine model does not claim that thermal conductance is always higher or always lower than the macroscopic expectation; rather, it shows that thermal conductance varies depending on molecular stiffness and friction. This is one of the key distinctions between our model, which defines thermal conductance at the scale of a single protein, and previous models [41], where thermal conductance is treated in the traditional bulk sense for the entire mitochondrion.

** The assumption of a thermally isolated axle does not suggest that analogous proteins in the IMM must be thermally insulated. Instead, it illustrates that even in this limiting case, heat transfer can still occur.

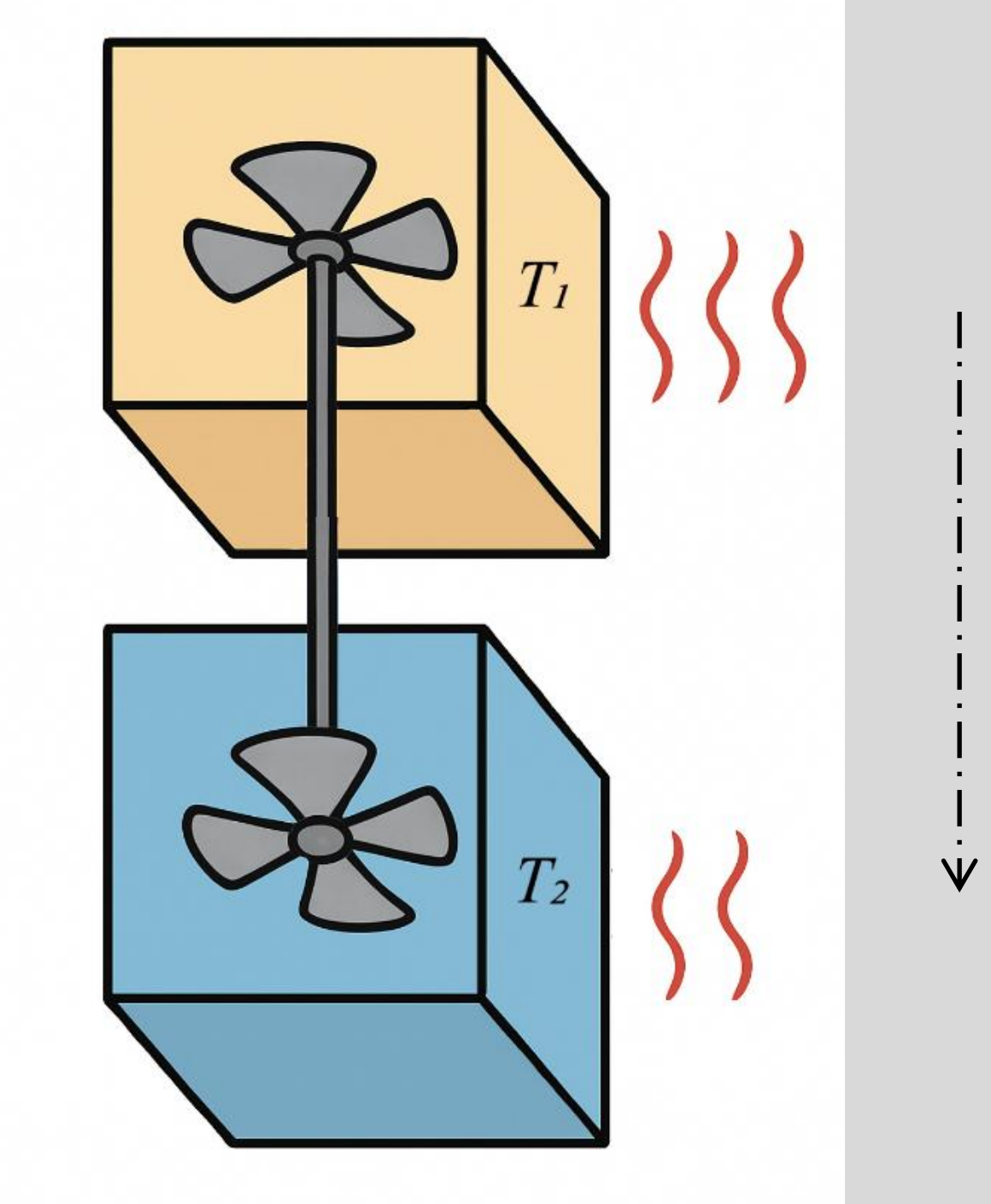

*Figure 1: The axle-vane engine, whose thermal conductance was originally formulated by Parrondo and Español [62] as a novel mode of heat transfer, and later used by us to estimate the thermal conductance of ATP synthase [9]. The engine consists of two vanes embedded in two thermally isolated boxes maintained at different temperatures ($T_1 > T_2$), connected by an insulated axle.*

In this section, we present the analogy between the axle-vane engine and ATP synthase, as introduced in our previous work [9]. In the following section, we extend this analogy to all other proteins embedded in the IMM. ATP synthase is the enzyme responsible for producing the majority of adenosine 5′-triphosphate (ATP) molecules in most living organisms during aerobic respiration. It is located in the inner mitochondrial membrane, the thylakoid membrane of chloroplasts, and the plasma membrane of bacteria. The enzyme consists of two main components: the Fo unit, which is embedded in the IMM, and the $F_1$ unit, which resides in the mitochondrial matrix (Figure 2 – left). Within the Fo unit, several subunits are involved, with the c-subunit and a-subunit playing key roles in capturing protons from the IMS and facilitating their translocation into the matrix (Figure 2 – left).

The significant pH difference across the IMM - acidic (high proton concentration) in the IMS and alkaline (low proton concentration) in the matrix - drives the binding of protons to various amino acid residues of the Fo unit, promoting forward proton translocation. Thus, the enzyme exhibits at least two forms of asymmetry: first, the proton concentration gradient, and second, the ratchet-like geometric asymmetry of the Fo and $F_1$ units. This unidirectional proton flow, combined with the unidirectional rotational motion of the $F_o$ unit (clockwise when viewed from the gap toward the matrix [63]), forms the basis of the analogy between the $F_o$ unit and a ratchet engine. This analogy has also been adopted by other researchers [52–56]. There is a trade-off between the rotational speed and efficiency of this enzyme when modeled as a ratchet engine - a relationship that has been analyzed by Wagoner and Dill from an evolutionary perspective [64].

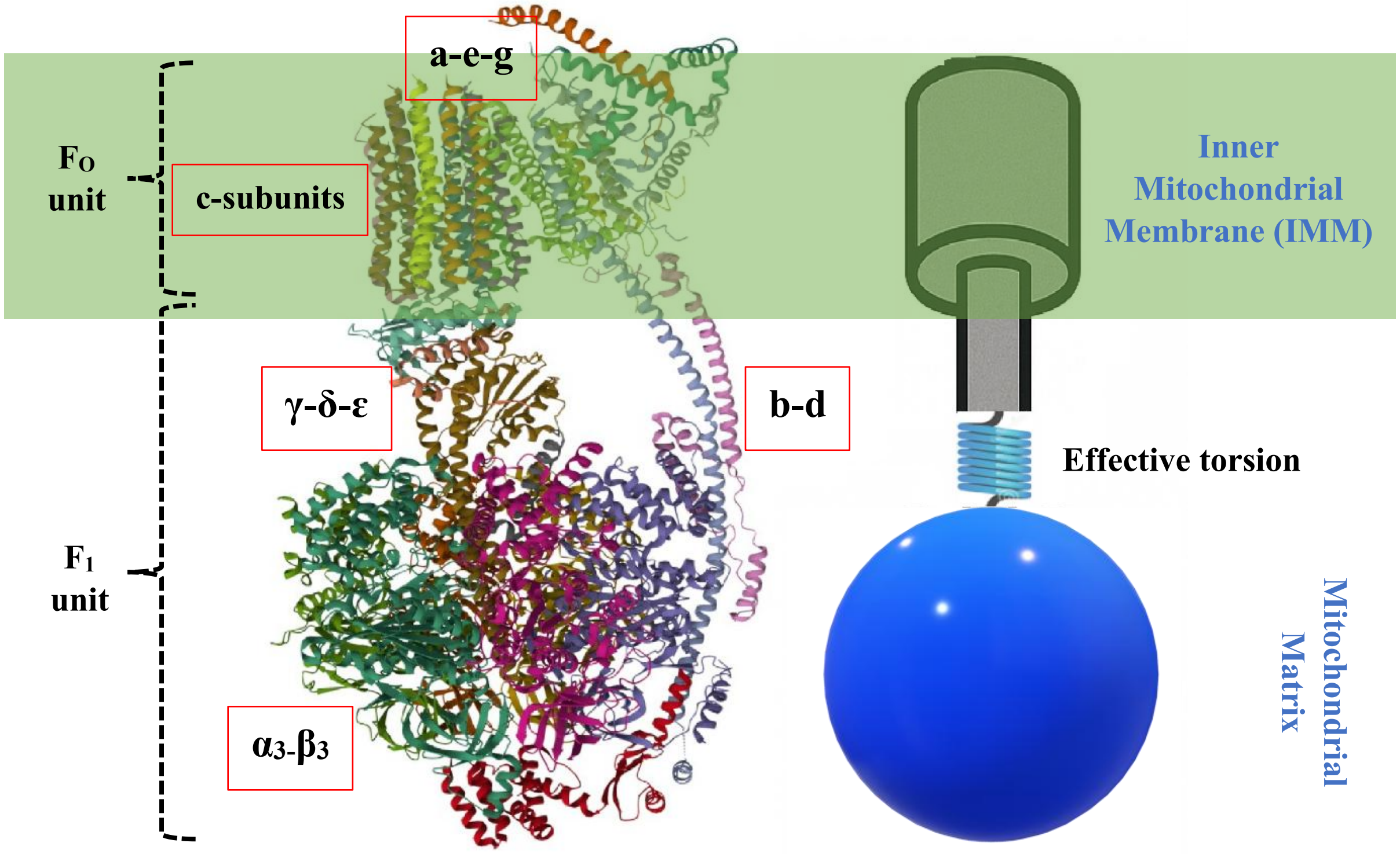

*Figure 2: The structure of human ATP synthase [65,66] containing eight c-subunits is shown in the left panel, with the Fo unit embedded in the inner mitochondrial membrane. The names of other major subunits are also indicated. A cartoon sketch on the right illustrates the modeled ATP synthase as a ratchet engine [9], where the γ-subunit is connected to the*

*$\alpha_3\beta_3$-subunit via an effective torsional spring. The size of the c-ring in ATP synthases varies widely across species, with experimental evidence showing between 8 and 17 subunits [67]. A recent study used an AlphaFold-based computational approach to predict the stoichiometry of homooligomeric c-rings from genomic data, suggesting that naturally occurring c-rings could contain up to 27 subunits - far exceeding the experimentally observed range [68]. However, in our model, the number of c-subunits has minimal impact on temperature gradients. This is because the effective spring illustrated here is torsionally soft, and even a three-fold increase in the moment of inertia of the c-ring connected to the γ-subunit has a negligible effect on temperature. For detailed estimations, see page 104 of our previous work [9].*

Stochastic differential equations are widely used to model systems subject to random fluctuations, such as Brownian motion. In this framework, the deterministic equation of motion - typically Newton's second law stating that force equals mass times acceleration - is modified by adding random noise terms to account for fluctuations. This results in the Langevin equation, which incorporates both deterministic dynamics and stochastic influences. When the ratchet-engine model of ATP synthase is represented using the Langevin equation, the random collisions of protons in the IMS with the $F_O$ unit's amino acid side chains serve as the noise generator or "stochastic forces" [9]. The electrostatic torque responsible for rotation is generated when a proton enters the channel of the $F_o$ unit [69].

The ratchet-engine mode of heat transfer in ATP synthase results in very low thermal conductance due to friction between the enzyme and its environment - primarily arising in the γ-subunit within the $F_1$ unit (Figure 2, left) - as well as due to the enzyme's torsional properties. In this model, the torsional stiffness is represented as an effective spring connecting the γ-subunit to the $\alpha_3\beta_3$ subunits (Figure 2, right), reflecting the torque required to rotate the enzyme by a given angle [9,70,71].

So far, we have discussed the thermal conductance of ATP synthase. We now turn our attention to the rate of heat production associated with proton translocation through the enzyme. Given that protons do not exist as such in aqueous media but that they rather exist as hydrated hydronium ions, and given that molecular-dynamic simulations suggest that only protons (not hydronium ions) are passed within ATP synthase's $F_o$ unit [72], for proton translocation to occur, the following process must be occurring (see also Figure 3). First, *endergonic processes occur on the IMS* since the hydration shell must first be stripped away from the hydronium ion followed by deprotonation of the hydronium ion to water and a proton. The proton is then captured by the amino-acid residues of the a-subunit or the a–c interface (the interface between the a-subunit and the c-subunit, Figure 3) of the $F_O$ unit, passing from one residue to the next within the channel until it exits on the matrix side of the IMM. The ejected proton in the mitochondrial matrix protonates a water molecule to form a hydronium ion, which is then solvated by the surrounding water, releasing the same amount of energy that was consumed on the other side of the membrane. At steady state and over relatively long timescales (e.g., milliseconds), the endergonic dehydration-deprotonation of a proton in the IMS is balanced by the exergonic protonation-hydration of the same proton in the matrix, such that the net heat production is effectively zero. However, over ultrashort timescales (1–100 ps, as discussed below), the protonation-hydration event and the associated heat released into the matrix can generate substantial transient temperature gradients [9]. The magnitude of these short-lived heat events is much greater than the steady-state bulk mitochondrial heat generated through proton leak, and this represents a major effect that has not been taken into account in previous studies.

As explained in Ref. [9], estimating the temperature difference across the IMM requires a knowledge of the protonation energy of water ($\Delta G^0_{\text{Protonation}}$), the hydration energy of the hydronium ion ($\Delta G^0_{\text{Hydration}}$), the rate of proton translocation through ATP synthase ($N_{\text{p.u.t.}}$, for number of proton translocated per unit time), the effective torsional stiffness of the axel of ATP synthase ($\tau_{\text{eff.}}$), and the friction between the enzyme subunits and their local environments during conformational changes ($\lambda$), as shown in the following equation developed in [9] (the full derivation of the equations used in [9] are provided in Chapter 1 of Ref. [73]):

$$\Delta T \cong \frac{2N_{\text{p.u.t.}}\lambda}{k_B \tau_{\text{eff.}}} \left( \Delta G^0_{\text{Protonation}} + \Delta G^0_{\text{Hydration}} + \Delta G^0_{\text{Heat bath}} \right) \qquad (1),$$

where $k_B$= 1.380649 × $10^{-23}$ J/K is the Boltzmann constant, and $\Delta G^0_{\text{Heat bath}}$ is the kinetic energy that the proton acquires from the thermal background of its surroundings upon translocation to the opposite side of the IMM. Upon binding to a water molecule, the proton transfers this kinetic energy to that molecule and its hydrogen-bonded neighbors through collisions. However, this energy is negligible compared to the free energy changes associated with protonation and hydration. *Equation (1) resembles Fourier's law but is formulated for a ratchet engine model, with parameter values specific to ATP synthase.* The term $\frac{2\lambda}{k_B \tau_{\text{eff.}}}$ corresponds to the thermal resistance or the inverse of thermal conductance, while $N_{\text{p.u.t.}} \times \left( \Delta G^0_{\text{Protonation}} + \Delta G^0_{\text{Hydration}} + \Delta G^0_{\text{Heat bath}} \right)$ denotes the heat power. The complete derivation is provided in Chapter 1 of the PhD dissertation [73].

Endergonic reactions at proton channel entry:

$\mathrm{H_3O^+(aq) + A(aq)}$
$\rightarrow \mathrm{H_3O^+(g) + A(aq)} + n\left(\mathrm{H_2O(aq)}\right)$
$\rightarrow \mathrm{H^+(g) + H_2O(g) + A(aq)} + n\left(\mathrm{H_2O(aq)}\right)$
$\rightarrow \mathrm{H_2O(aq) + AH^+(aq)}$

Exergonic reactions at proton channel exit:

$\mathrm{H_2O(aq) + AH^+(aq)}$
$\rightarrow \mathrm{H^+(g) + H_2O(g) + A(aq)} + n\left(\mathrm{H_2O(aq)}\right)$
$\rightarrow \mathrm{H_3O^+(g) + A(aq)} + n\left(\mathrm{H_2O(aq)}\right)$
$\rightarrow \mathrm{H_3O^+(aq) + A(aq)}$

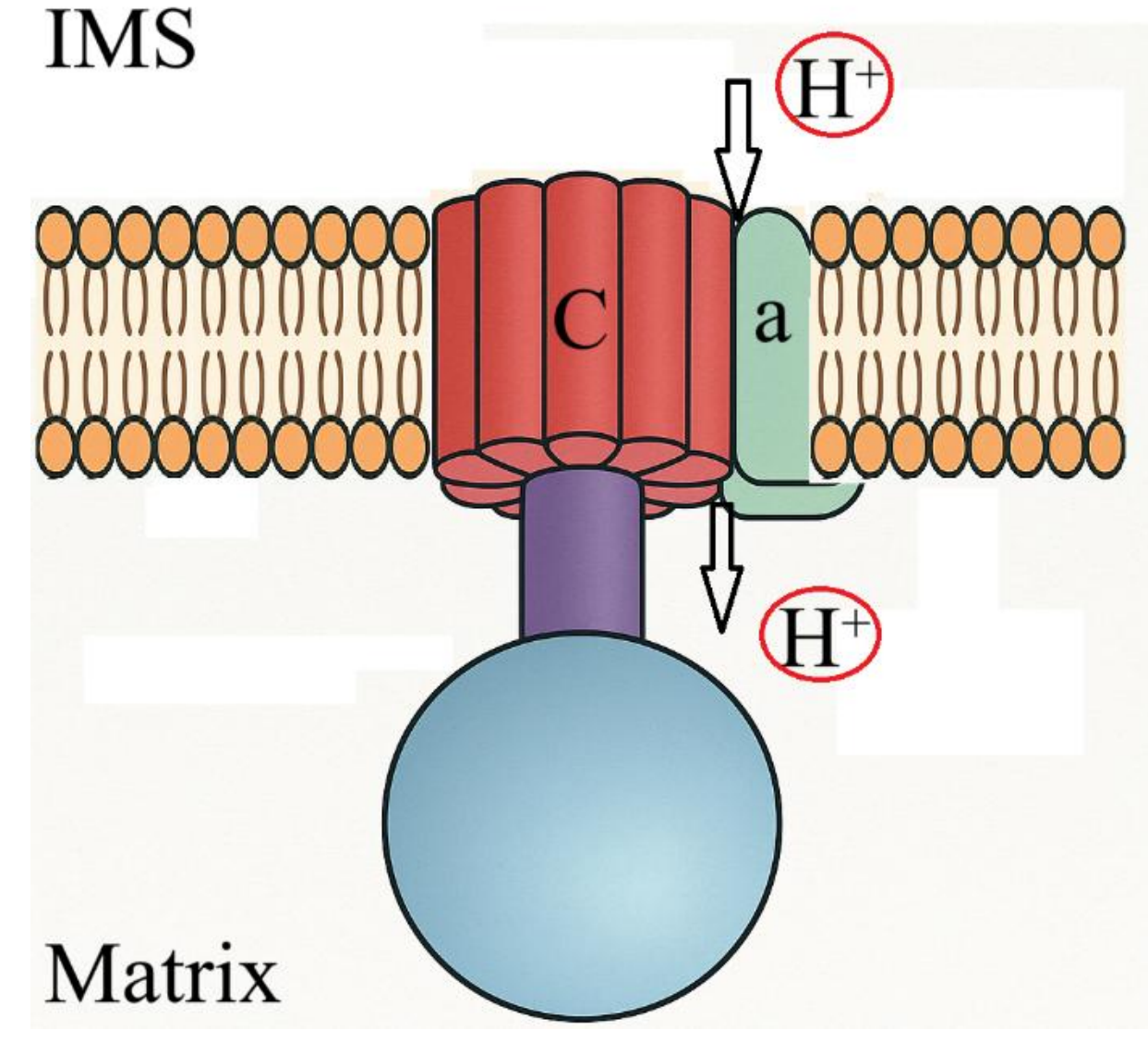

*Figure 3: The endergonic and exergonic reactions occurring at the entry and at the exit of the proton channel, respectively. (**Left**) $H_3O^+(aq)$ represents the hydronium ion in the aqueous phase, A(aq) denotes the first amino acid residue that captures the proton, (g) refers to the gas phase, $n(H_2O(aq))$ indicates the number of water molecules released during dehydration, and $AH^+(aq)$ is the final product of the reaction, showing that the proton has been successfully captured by the amino acid. (**Right**) A representation of this process, where protons enter through the interface between the c-subunit and a-subunit in the intermembrane space (IMS) and are subsequently released into the mitochondrial matrix after translocation.*

Note that we sum over all Gibbs free energy changes in Equation (1) because they occur within a few picoseconds - faster than the timescale of heat diffusion - which justifies treating them collectively. In other words, after analysis of the timescales of the various processes (protonation, and hydration) and the thermalization, the heat imbalance is estimated to thermalize within a few picoseconds as discussed in detail in our previous work [9]. However, we emphasize that several ultrafast studies show that local vibrational energy in proteins can persist for 10–100 ps [74–76], indicating that heat-diffusion timescales can, under certain conditions, be even longer. This occurs because: (1) many of the vibrational modes of proteins are spatially localized, slowing energy propagation [74]; (2) heat must flow through specific atomic-contact pathways, creating bottlenecks [75]; and (3) part of the relaxation involves slow, low-frequency protein motions [76]. These effects have been observed in (a) theoretical analyses of myoglobin showing 1–10 ps mode lifetimes and 20–30 ps protein to solvent heat flow due to mode localization [74]; (b) myoglobin, where vibrationally excited residues remain hot for up to 50 ps [75]; and (c) flavoenzymes, where FMN to protein energy transfer shows components up to ~100 ps [76].

The combined energy of water protonation and subsequent hydronium ion hydration amounts to approximately 1,200 kJ/mol ($\approx 2.00\times10^{-18}$ J/ $H^+$). The rate of proton translocation is $N_{\text{p.u.t.}} \approx 1{,}200$ $H^+$/s, the friction coefficient between ATP synthase rotatory mechanism and its environment $\lambda \approx \frac{k_B T}{D} = \frac{4.11\times10^{-21}\text{J}}{0.001\times10^{9}\frac{\text{rad}^2}{\text{s}}} = 4.11 \times 10^{-27}\text{J.s/rad}^2$ [9] ($\text{rad}^2$ is the square of angular displacement), and the effective torsional stiffness of the axel is $\tau_{\text{eff.}} \approx 223 \times 10^{-21}\text{J/rad}^2$ [71]. By inserting these parameters into Equation (1) [9], a short-lived temperature difference of approximately 6°C–7°C is estimated [9]. Since the parameter values in Eq. (1) vary across physiological conditions and experimental techniques, we performed a sensitivity analysis using literature-reported ranges for ATP synthase. As shown in the Supporting Information, the resulting temperature spikes span approximately 2–37 °C, with a mean of about 6.5 °C.

## 3. Can IMM proteins other than ATP synthase be modeled as a ratchet engine?

In this section, we explore whether proteins of the inner mitochondrial membrane (IMM), aside from ATP synthase, can also be modeled as ratchet engines. Proteins within the IMM - such as ion exchangers, mitochondrial carrier proteins (MCPs), uncoupling proteins (UCPs), ABC transporters, and respiratory complexes (for a comprehensive review of ion channels and SLC25 carrier family in the mitochondrial membrane, see references [77–80])

- mostly share a common mechanism of energy conversion and transport that is consistent with the behavior of ratchet engines. These proteins exploit electrochemical gradients of charged species or substrate exchange gradients or redox potential differences to facilitate unidirectional transport, driven by random collisions and conformational changes under non-equilibrium conditions. The modeling of ion pumps (such as $Na^+/K^+$-ATPase [59] and the bacteriorhodopsin proton pump [57,58]) and voltage-dependent ion channels [60] as ratchet engines has been extensively explored in the literature.

In physical models of ratchet engines [53,61] and axle-vane engines [62], the random Brownian motion of particles is assumed to directly drive engine rotation - for example, the rotation of the c-subunits in ATP synthase as described in the previous section. Here, we extend the definition of physical ratchet engines to include other IMM proteins. Ions and metabolites undergoing Brownian motion frequently collide with these proteins due to electrochemical potential or redox potential differences. When an ion is recognized and captured by a protein, the protein can harness this interaction to perform work, often manifested as "*conformational changes*". These conformational changes, in turn, drive the protein's functional or mechanical activity facilitating the translocation of the ion.

Consider the three-dimensional IMM as being composed of an infinite number of segments. Each segment maintains an electrochemical ion gradient (either from the matrix to the IMS or *vice versa*) across the membrane or a redox potential difference. For instance, because of the large redox potential drop along the electron transport chain, NADH and $FADH_2$ reach their respective respiratory complexes by Brownian diffusion and bind to high-affinity redox centers. Upon binding, electrons flow spontaneously down their redox potentials, releasing Gibbs free energy that the complexes harness to capture and pump protons from the matrix to the intermembrane space. Simultaneously, other ions such as $Na^+$ and $Ca^{2+}$ undergo exchange across the membrane as dictated by their electrochemical gradients and physiological requirements. Proteins of the IMM function unidirectionally *for a given time and a given ion gradient or redox potential difference* consistent with the ratchet-engine model. One might question whether ion exchangers, such as the sodium-proton exchanger in an antiport system, exchange ions simultaneously, which would seem to contradict unidirectionality. However, even in this type of channel, the exchange does not occur exactly at the same time, as the protein cannot undergo two different conformational changes simultaneously. Instead, the processes occur in succession, *making each individual event unidirectional in time*.

The conformational changes that facilitate ion transfer are influenced by factors such as the friction between the moving protein subunits and their surrounding environment, as well as the torsional or linear stiffness of the protein. Protein stiffness depends on the mechanical properties of the protein, particularly the stiffness of the subunit undergoing conformational changes that facilitate ion translocation. Additionally, the friction between protein subunits and their membrane-bound environment varies with several factors, including the membrane electrochemical potential (pH and electric field), temperature, lipid composition, and hydrodynamic, hydrophobic, and steric interactions, all of which can differ from one protein to another. It is also important to note that the micro-viscosity of the IMM is not constant; it varies with the electric field strength of the membrane, which is influenced by the cell's energy status - whether satiated or starved [81]. Thus, an actively growing cell with a higher proton motive force across the IMM generates a stronger electric field, which in turn increases the IMM's viscosity and friction [81].

When the electrochemical gradient of a specific ion per segment at a given time is higher on one side, the collision frequency of those ions with the corresponding protein increases. In the case of respiratory complexes I, III, and IV, a similar sequence of events occurs: a high redox potential drives random collisions between electron donors and the complexes, while hydronium ions also collide with them. The complexes then use the ΔG provided by the electron-transport chain to capture protons and pump them into the IMS. In terms of the ratchet model, *an ion gradient is analogous to a temperature gradient* (meaning that more random Brownian motion per segment → more kinetic energy → higher temperature), as it drives directional work through stochastic processes. For ions to be translocated across membranes via channels and pumps, they are generally (at least partially) dehydrated in order to be "*felt*" by the channel [82]. This is well-documented in the cases of sodium and potassium channels [83,84]. An exception occurs when the pore is sufficiently wide to allow ions to pass through with their hydrated shells intact [85]. Once an ion is dehydrated, it is captured by amino-acid residues within the protein channel through random Brownian collisions. These interactions induce conformational changes in the protein, facilitating the ion's translocation to the other side of the membrane, where it rehydrates with water molecules. The emerging view is that temperature spikes occur on both sides of the IMM (i.e., within both the IMS and the matrix), making the combined "IMS + matrix" region hotter than the cytoplasm. Importantly, the fact that both the IMS and the matrix are hotter than the cytoplasm does not imply that they maintain the same steady temperature. The system is fundamentally non-equilibrium, comprising an effectively continuous set of micro-domains, each of which may be hotter, but at different temperatures at any moment.

Applying Eq. (1) to IMM proteins other than ATP synthase requires experimental parameter values for each protein type. To the best of our knowledge, $\tau_{\mathrm{eff.}}$ and $\lambda$ have not been reported for any mitochondrial ion-translocating proteins aside from ATP synthase. However, we have compiled $N_{\mathrm{p.u.t.}}$ and the $\Delta G$ of hydration for a set of major ions and their corresponding mitochondrial proteins, as summarized in Table 1.

The energy of hydration varies for different ions and metabolites, depending on their charge, size, and chemical properties and is crucial in determining the temperature difference across the IMM (see Eqn. (1) above and details in Ref. [9]). Given the role of hydration (and dehydration) in the transport of ions through biological membranes, in Table 1 we list known approximate Gibbs energies of hydrations at standard temperature (298.15 K) and pressure (1 atm) for common ions and metabolites encountered in cellular processes together with their rate of translocation per second. Unlike the protonation energy, which is a well-defined chemical reaction where one hydrogen ion ($H^+$) combines with a water molecule ($H_2O$) to form a hydronium ion ($H_3O^+$) in a one-to-one ratio ($H_2O + H^+ \rightarrow H_3O^+$), hydration doesn't follow a fixed formula. This is because hydration refers to how water molecules surround and interact with a solute (like an ion or molecule) in solution. These water molecules form layers (called solvation shells) around the solute, but the exact number and arrangement of water molecules are not fixed. They change constantly due to the motion and interactions in the liquid, making hydration a statistical and dynamic process as captured for instance in molecular-dynamics simulations where it is typical to plot distribution functions of solvent molecules around the solute molecule or ion [86,87].

Table 1: Gibbs free energies of hydration for selected ions, along with their rates of translocation through the corresponding membrane proteins in a cellular environment.

| Ion | $\Delta G$(hydration), kJ/mol | Ref. | Rate of ion translocation $s^{-1}$ | | Ref. |
|---|---|---|---|---|---|
| $H^+$ | ~ -1183 (a) | [88] | ATP synthase | 800 – 3500 | [89,90] |
| | | | Complex I (b) | 400 – 1640 | [91,92] |
| | | | Complex III (b) | 20 – 1632 | [93] |
| | | | Complex IV (b) | 600 | [94] |
| | | | UCP1 (c) | 3 – 1000 | [95] |
| $Na^+$ | -406 | [96] | $Na^+$/ $Ca^{2+}$ exchanger (NCLX) (d) | 3000 | [97] |
| $K^+$ | -322 | | Potassium channel | $10^8$ | [98] |
| $Ca^{2+}$ | -1577 | | Calcium uniporter (MCU) | $5.0\times10^6$ | [99] |
| $Mg^{2+}$ | -1921 | | $Mg^{2+}$ transporter (Mrs2p) (e) | $9.7\times10^7$ | [100] |
| $Cl^-$ | -363 | | Chloride channel (e) | $(2.0 – 3.8)\times10^8$ | [101] |
| $H_2PO_4^-$ | -337 | [102] | Inorganic phosphate carrier (PiC) (f) | 800 – 1000 | [103] |
| $ATP^{4-}$ | -2754 | | Adenine nucleotide translocase (ANT) (g) | 5 – 100 | [104] |
| $ADP^{3-}$ | -1799 | | | | |

(a) This value is for the sum of Gibbs energies of protonation and of hydration.
(b) For Complexes I, III, and IV, the literature turnover numbers were scaled by four, assuming four protons are pumped per catalytic cycle, to obtain the total proton-translocation rate per second.
(c) For UCP1, the turnover number equals the proton-translocation rate.
(d) For NCLX, a turnover rate of 1000 $s^{-1}$ gives 3000 $Na^+$ and 1000 $Ca^{2+}$ exchanged per second, based on a 3:1 $Na^+$/$Ca^{2+}$ stoichiometry.
(e) For Mrs2p and the chloride channel, ion-flux rates are estimated from channel conductance.
(f) For PiC, a turnover of 50,000–60,000 $min^{-1}$ (833–1000 $s^{-1}$) is reported, with each cycle transporting one Pi and one $H^+$.
(g) For ANT, the exchange rate is 5–100 $ATP^{4-}$ for 5–100 $ADP^{3-}$ per second.

The rate of ion translocation in ion channels, such as potassium channels (Table 1), is ~ $10^8$ ions/s [98], five orders of magnitude higher than the rate of proton translocation through ATP synthase. This raises the question of whether it results in a temperature difference five orders of magnitude greater than the one we estimated for ATP synthase. The answer is no, primarily because the rate of translocation depends on the friction between the enzyme and its surrounding membrane lipids. The γ-subunit diffusion coefficient (≈0.001 $rad^2$/ns [70]) is about 70 times lower than in solution, indicating markedly higher friction within ATP synthase [105]. The friction opposing conformational changes in an ion channel must be significantly lower than that involved in the full rotation of ATP synthase, allowing the ion channel to support a much higher rate of ion translocation. Thus, the high rate of transport by ion channels is cancelled out by the extremely low friction against their conformational changes (see Eqn. (1)), resulting in temperatures similar to those we estimated for ATP synthase. Thus, depending on the parameter values in the term $\frac{2\lambda}{k_B\tau_{\text{eff.}}}$ in Eq. (1), ATP synthase acts as a relatively poor thermal conductor, whereas potassium channels function as relatively efficient thermal conductors. Although we currently lack direct measurements of the torsional or linear stiffness of ion channels and the friction opposing their conformational changes, the theoretical hydrodynamic model developed by Chen *et al.* [106], which is distinct from ours, predicts that the temperature of sodium ions in channels with dimensions resembling gramicidin A (25 Å in length and 2 Å in radius) can increase by approximately 25 °C under a 100 mV transmembrane potential, with the magnitude of this increase depending on the applied electrical field and the saturation velocity of ions in the channel.

The overarching picture that emerges is that any membrane - whether of an organelle or a whole cell - that is highly selective and contains a high protein-to-lipid ratio can potentially become "hot", provided that the combined effects of ion-translocation rates ($N_{\text{p.u.t.}}$), the $\Delta G$ of hydration events, protein torsional or linear stiffness ($\tau_{\text{eff.}}$), and friction ($\lambda$) between protein subunits and their surrounding environment generate a substantial temperature gradient. Table 1 lists several major ions translocating across the IMM, along with their $\Delta G$ of hydration, and translocation rates per second, offering a basis for estimating the potential contribution of each protein to temperature-spike generation. However, to the best of our knowledge, the parameters $\tau_{\text{eff}}$ and $\lambda$ remain unknown for the proteins (other than ATP synthase) listed in Table 1, representing an important avenue for future investigation.

## 4. Interpreting the Signals Recorded by Molecular Probes

The crucial question is: *what are molecular probes actually measuring?* Are millisecond-scale, membrane-wide signals detected by these probes simply synchronized manifestations of underlying picosecond thermal spikes? Or are the probes capable of detecting the picosecond temperature fluctuations themselves? In the following, we examine both possibilities. Potential experimental strategies and falsifiable predictions that could test this framework are summarized in Box 5.

*Synchronized temperature spikes in highly active mitochondria*: If ion translocations by one protein are staggered with those of neighboring ones, temperature difference across the IMM would be continuous over time in this neighborhood. How can such synchronization potentially occur in the IMM?

When the mitochondrion operates at maximum capacity, the cristae membranes become tightly packed [107,108] and densely populated with proteins, creating *confined spaces* that may retain heat locally for periods longer than a few picoseconds. As discussed in the final paragraphs of Section 2, localized vibrational modes and bottlenecked energy pathways can extend protein heat-diffusion lifetimes to 10–100 ps.

An ensemble of all proteins of the IMM would have ion translocation events at a rate $\mathbf{P} = \sum_{i=1}^{n} P_i \approx \sum_{i=1}^{n} \dot{N}_i M_i$ where $\dot{N}$ is the number of ions translocated per second per protein type (e.g. $P_1$), $M$ is the copy number of such protein molecules in the neighborhood sensed by the fluorescent probe, and $n$ is the number of protein types ($P_1, P_2, P_3, \dots, P_n$). We provide two examples here. The number of active ATP synthases depends primarily on the cellular metabolic rate and the mitochondrial population. As we estimated previously for heterotrophic protists [44], highly active organisms can have more than 30,000 ATP synthases per mitochondrion (refer to the Supporting Information for an explanation of how this estimate was obtained.). Thus, the number of proton translocation events (assuming 1200 $H^+$ per ATP-synthase per second) would be $P_1$ = 1200 × 30,000 = 3.6 × $10^7$. Assuming each temperature spike lasts 1 to 100 ps as discussed in the previous paragraph, we obtain a cumulative thermal occupancy of $\Theta$ = 3.6 × $10^7$ $s^{-1}$ × (1 to 100) × $10^{-12}$ s = 0.036 to 3.6 milliseconds per second. As a second example, an intermyofibrillar rat cardiac mitochondrion with dimensions of approximately 1.5 µm × 0.3 µm × 0.3 µm has a total IMM surface area of about 5.76 µm² [109]. The mitochondrial calcium uniporter (MCU) in rat heart IMM has an estimated channel density of 10–40 channels per µm², and each channel supports a maximum $Ca^{2+}$ flux of ~5 × $10^6$ ions $s^{-1}$ [99]. Thus, the total $Ca^{2+}$ translocation events per mitochondrion would be 5 × $10^6$ ions $s^{-1}$ × 40 channels $\mu m^{-2}$ × 5.76 µm² ≈ 1.15 × $10^9$ ions $s^{-1}$. If each temperature spike lasts 1 to 100 ps, then the cumulative thermal occupancy of such events is $\Theta$ = 1.15 × $10^9$ $s^{-1}$ × (1 to 100) × $10^{-12}$ s = 1.15 to 115 milliseconds per second. Here we considered only two protein types ($P_1$ and $P_2$). Now imagine extending this to $n$ distinct IMM proteins (which would require knowing the population density of all protein types in the IMM when the mitochondrion is operating at its maximum respiratory capacity): the total number of ion-translocation events per mitochondrion becomes extraordinarily large.

This is not the end of the story - coupled synchronization of temperature spikes among these proteins may further amplify the effects described above. ATP synthases and other ion translocators can be regarded as oscillators, with their frequencies determined by the rates of ion translocation. In the IMM, there are at least two potential *modes of coupling* between neighboring proteins: local and global. Local coupling occurs when a temperature spike generated by one protein transiently accelerates reactions in nearby proteins according to Arrhenius kinetics. Global coupling arises through the proton motive force (PMF) across the IMM. Each ion translocation event perturbs the PMF, and this perturbation can in principle be sensed by other proteins embedded in the membrane. Consequently, proteins that are physically distant may still influence one another through a shared electrochemical environment. If such coupling promotes synchronization, thermal spikes that would otherwise occur randomly in time may become increasingly clustered, increasing the temporal overlap of heat-release events. This behavior may be described using Kuramoto-like models of coupled oscillators [110,111]. Whether such synchronization occurs in vivo, and whether it significantly influences mitochondrial temperature, remains an important subject for future investigation.

A useful way to interpret the total cumulative thermal occupancy is through the dimensionless quantity $\Theta_{\text{total}} = \sum_{i=1}^{n} P_i \tau_i$, where $P_i = \dot{N}_i M_i$ is the total flux associated with protein type $i$ per mitochondrion, as defined earlier, and $\tau_i$ is the lifetime of the corresponding thermal spike. Physically, $\Theta_{\text{total}}$ represents the expected number of thermal spikes simultaneously active on the IMM at any given instant. Thus, $\Theta_{\text{total}} \ll 1$ corresponds to a regime in which thermal spikes are sparse and largely isolated, whereas $\Theta_{\text{total}} \approx 1$ marks the onset of significant temporal overlap. When $\Theta_{\text{total}} > 1$, multiple thermal spikes are expected to coexist simultaneously, and for $\Theta_{\text{total}} \gg 1$ the membrane enters a strongly overlapping regime in which transient thermal events become densely distributed in space and time. Importantly, $\Theta_{\text{total}}$ should not be interpreted as a probability of overlap, but rather as a control parameter quantifying the degree of collective thermal activity on the IMM. In highly active mitochondria, where many classes of ion-translocating proteins operate concurrently, the cumulative thermal occupancy may substantially exceed unity, suggesting that overlapping thermal perturbations could become a common feature of the membrane environment.

ATP synthase and MCU discussed above are presented as illustrative examples to estimate the cumulative thermal occupancy associated with individual protein types. They are not intended to imply that all ATP synthases or MCUs are clustered around or attached to a single probe. These examples demonstrate that even one or two protein types can contribute to the cumulative thermal occupancy, despite that mitochondria contain additional ion-translocating proteins whose contributions have not been included. These estimates do not account for contributions from protein clustering, coupling-induced synchronization, or potentially longer effective thermal diffusion lifetimes. Consequently, the values calculated here are lower-bound estimates of the total cumulative thermal occupancy. Whether the combined activity of all IMM proteins can drive $\Theta_{\text{total}}$ toward or beyond unity *in vivo* remains unknown and constitutes a testable prediction of this hypothesis.

*Unsynchronized temperature spikes in less active mitochondria*: Imagine mitochondria that are not highly active: the number density of cristae membranes and IMM proteins is low, the cumulative rate of ion translocation is reduced, and there is little or no synchronization among temperature spikes. Under these conditions, can fluorescent probes still detect such spikes? Experimental fluorescence thermometry is limited to millisecond-scale resolution [112,113] which reflects instruments constraints and *not necessarily* the intrinsic response time of fluorescent probes. It remains an open question whether these molecular thermometers can sense rapid, localized temperature spikes on picosecond to nanosecond scales. On one hand, quantum chemical calculations offer a path to investigate this possibility. *Ab initio* (and post-Hartree-Fock) calculations [114–117], density functional theory (DFT) [118–120], and time-dependent DFT (TD-DFT) [118–120] can model the photophysical properties of fluorescent dyes' response to sudden thermal perturbations. TD-DFT, for example, can predict how excitation energies and oscillator strengths shift due to temperature-induced structural changes occurring within the nanosecond timescale of the dye's excited-state lifetime. On another hand, quantum molecular dynamics (e.g., Car-Parrinello MD simulations [121,122]) can probe the spatial range of thermal effects, revealing how hydration shells reorganize around the dye in response to localized heating. While the experimental signal is presumed to reflect an integral of the signals from many probes over time and space, the underlying emission could still reflect these nanoscale thermal spikes.

**BOX 5. Experimental strategies and falsifiable predictions for the proposed model**

Although a full experimental investigation is beyond the scope of this theoretical perspective, we briefly outline how the framework proposed here can, in principle, be tested or falsified using existing technologies. Several reported artifacts in intracellular thermometry arise from pH, viscosity, ionic strength, dye aggregation, membrane potential, or local electric fields. Published calibration data for MTY and related probes already quantify many of these effects (see Box 2); however, the present framework makes distinct predictions that can be controlled using (1) simultaneous pH- and temperature-responsive dyes to confirm that the observed quenching cannot be reproduced by physiologically reasonable pH shifts, (2) calibration against nanodiamond thermometry in the same cell type, which is insensitive to local chemical environment but sensitive to electric-field artifacts, and (3) chemical inhibition of ion-translocation steps (e.g., oligomycin for ATP synthase [123], DS16570511 for MCU [124], bongkrekic acid for ANT [125]). The model explicitly predicts that suppressing ion translocation should proportionally suppress temperature-spike-driven signals. Best practice therefore requires parallel calibration of molecular thermometers such as MTY against environment-insensitive tools like nanodiamond thermometry, ensuring that fluorescence changes reflect genuine thermal effects rather than local chemical or electrical perturbations.

The ratchet-engine / hydration-dehydration mechanism leads to several experimentally testable predictions. Purified IMM proteins (ATP synthase, ANT, MCU, UCP1, etc.) reconstituted into liposomes with well-controlled electrochemical gradients should exhibit protein-specific temperature spikes, which could be detected using fast infrared probes or nanodiamond thermometry, although it should be noted that liposomes possess a single membrane rather than the double-membrane architecture of mitochondria. Detecting such effects would require thermometric readouts with millisecond or sub-millisecond temporal resolution and sensitivity comparable to that reported for state-of-the-art fluorescent probes and nanodiamond-based sensors. Replacing transported ions with analogs of known hydration energies should change the magnitude of the thermal response in proportion to $\Delta G$ predicted by Eq. (1). The cumulative signal detected by the fluorescent probes should scale with the total flux $\mathbf{P}$, as predicted in Section 4, which can be approximated by $\sum_{i=1}^{n} \boldsymbol{P_i} = \sum_{i=1}^{n} \dot{\boldsymbol{N}}_i \boldsymbol{M_i}$, where $\boldsymbol{P_i}$ denotes a specific protein type (e.g., ATP synthase, calcium uniporter, complex I, etc.), $\dot{\boldsymbol{N}}_i$ is the ion translocation per second per molecule of protein $\boldsymbol{P_i}$, $\boldsymbol{M_i}$ is the copy number of that protein sensed by the probe, and $\boldsymbol{n}$ is the total number of protein types contributing to the signal. Mutations that alter torsional or linear stiffness or friction in the protein (e.g. ATP synthase) are predicted to alter $\Delta T$ in a direction consistent with Eq. (1).

## Closing Remarks

This article revisits the "*Hot-Mitochondrion Paradox*" (HMP), the apparent contradiction between experimental evidence suggesting mitochondria operate ~10-15°C hotter than their environment and theoretical limits imposed by classical steady-state heat transfer as stipulated by Fourier's law. The fluorescent thermometry results of Chrétien *et al.* suggest a five to six -order-of-magnitude discrepancy from steady-state Fourier-law prediction of the temperature difference across the IMM, known in the literature as the "$10^5$ gap".

The proposed resolution of the hot-mitochondrion paradox rests on two complementary ideas: (i) ion translocation events can generate transient localized temperature spikes that are not captured by steady-state analyses, and (ii) membrane proteins may exhibit microscopic thermal conductances that differ substantially from bulk expectations when described within a ratchet-engine framework. Together, these effects offer a possible explanation for the long-standing discrepancy between intracellular thermometry measurements and classical heat-transfer estimates.

We propose resolving this paradox by modeling IMM proteins as Brownian ratchet engines that generate short-lived (1–100 ps) temperature spikes during proton or ion translocation, ranging from ~2°C to ~37°C with a mean of ~6.5°C (see Supporting Information), consistent with experimental fluorescence thermometry. We further propose that the collective activity of numerous ion-transporting membrane proteins can generate sustained local temperature differences in the vicinity of a fluorescent dye molecule, detectable within the temporal resolution of fluorescent thermometry. However, the spatial distribution of these temperature spikes remains unresolved: whether they are confined to a narrow layer near the inner mitochondrial membrane or extend throughout the entire mitochondrion is a question for future investigation.

The framework developed in here is generalizable to other membranes with a high protein-to-lipid ratio and implies broader implications for thermally mediated cellular processes. The ratchet model coupled with the dehydration-translocation-rehydration sequence of events and the presence of a large density of proteins thus offers a physically plausible and mechanistically sound reconciliation of theory with observation.

If mitochondrial membrane proteins generate short-lived temperature spikes, is this something evolution has adapted for (to speed up reactions) or is it merely an unavoidable byproduct of biological processes? It's likely an inevitable byproduct, but evolution may have taken advantage of it where useful. Think of it like a car engine: the engine gets hot as a side effect of running. Engineers might place heat-sensitive components near the engine to make use of that heat. But the heat isn't the goal, it's a byproduct that can sometimes be useful. Future experiments could test this hypothesis by determining whether mitochondrial membrane proteins located near regions of high ion-translocation activity exhibit systematically enhanced reaction kinetics compared with proteins in less active regions. This hypothesis could also be tested in bacterial membranes, particularly in *Vibrio natriegens*, one of the fastest-growing bacteria known [126] and a species characterized by exceptionally high substrate uptake rates [127,128]. If transmembrane ion fluxes generate transient, localized temperature fluctuations analogous to those proposed for mitochondrial membrane proteins, then the exceptionally high ion fluxes of *V. natriegens* may produce larger thermal fluctuations that enhance membrane-enzyme kinetics and potentially contribute to its remarkable growth rate.

It is fitting to conclude with a quote from Sainsbury's 1990 book on Paradoxes:

> "This is what I understand by a paradox: an apparently unacceptable

> conclusion derived by apparently acceptable reasoning from apparently acceptable premises. Appearances have to deceive, since the acceptable cannot lead by acceptable steps to the unacceptable. So, generally, we have a choice: either the conclusion is not really unacceptable, or else the starting point, or the reasoning, has some non-obvious flaw.
>
> Paradoxes come in degrees, depending on how well appearance camouflages reality. Let us pretend that we can represent how paradoxical something is on a ten-point scale. The weak or shallow end we shall label 1; the cataclysmic end, home of paradoxes that send seismic shudders through a wide region of thought, we shall label 10".

According to this scale, we imagine that the hot mitochondrion paradox, where theory and experiment disagree neither by a factor of 2 factor nor by a factor of 100, but rather by a factor of one million deserves a place at the summit of the pyramid of paradoxes with a well-deserved label of 10.

## Acknowledgments

The authors thank Professor Thanh-Tung Nguyen-Dang (Laval University) for his helpful discussions, as well as the editor and four anonymous reviewers for their constructive comments. CM is grateful to the Natural Sciences and Engineering Research Council of Canada (NSERC), the Canada Foundation for Innovation (CFI), Saint Mary's University, Dalhousie University, Mount Saint Vincent University, and the Digital Research Alliance of Canada for their financial support and resources. ML is grateful to the National Institutes of Health (2R35GM122566), the National Science Foundation (DBI-2119963, DEB-1927159), the Simons Foundation (735927), and the Moore Foundation (12186).

## Data availability statement

All data supporting the findings of this study are available within the article and its Supporting Information.